# Optimization Studies of Radiation Shielding for the PIP-II Project at Fermilab

*Alajos* Makovec [1*], *Dali* Georgobiani[1], *Igor* Rakhno[1], and *Igor* Tropin[1]      FERMILAB-CONF-25-0355-AD-PIP2

[1]Fermi Forward Discovery Group, LLC (FermiForward), Fermi National Accelerator Laboratory (Fermilab), Batavia IL 60510, USA

**Abstract.** The PIP-II project at Fermilab, which includes an 800-MeV superconducting LINAC, demands rigorous radiation shielding optimization to meet safety requirements. We updated the MARS geometry model to reflect new magnet and collimator designs and introduced high-resolution detector planes to better capture radiation field distributions. To overcome the significant computational demands, we implemented a well-known branching technique that drastically reduced simulation runtimes while maintaining statistical integrity. This was achieved through particle splitting and the application of Russian Roulette techniques. Additionally, new graphical tools were created to streamline data visualization and MARS code usability.

## 1 Introduction

The Proton Improvement Plan II, or PIP-II, is a crucial upgrade to the Fermilab accelerator complex, designed to produce the world's most intense high-energy neutrino beam. This enhancement features a new 800 MeV superconducting radio-frequency linear accelerator, which will enable the Fermilab complex to deliver more than a megawatt of beam power to the Long-Baseline Neutrino Facility (LBNF). The project is scheduled for completion in 2028.

To ensure safety and regulatory compliance, it is essential to assess dose rates under both normal operational and accidental scenarios, in accordance with the Fermilab Radiological Control Manual (FRCM). To accurately reflect the current state of the PIP-II structure, we updated the existing MARS geometry model to include newly designed magnets and collimators. Radiation transport simulations were performed using the Monte Carlo code MARS [1, 2, 3] to support shielding optimization.

This work builds on prior model developments and simulation studies related to PIP-II, the Beam Transfer Line (BTL), and the collimation system, previously carried out by members of the MARS group: I. Tropin, I. Rakhno, and D. Georgobiani.

## 2 Implementation of Parametric Geometry

To ensure that the MARS geometry remains a reliable and comprehensive tool for the PIP-II project (see Fig. 1), several key components needed to be integrated: two collimators (horizontal and vertical PIP-II collimators) and two magnets (Fig. 2). This chapter focuses on the implementation of the collimators.

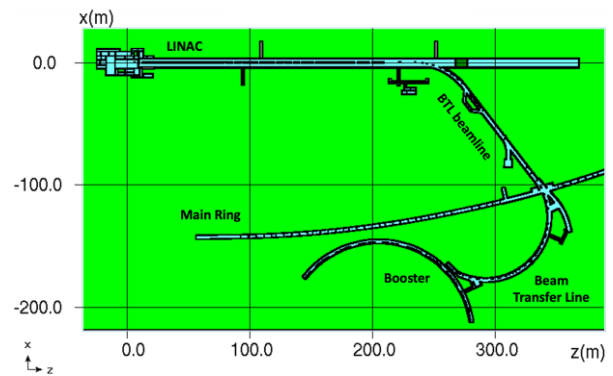

**Fig. 1.** Top view of the MARS geometry layout, showing the linear accelerator (LINAC) at the top and the BTL beamline extending to the right.

Both the horizontal and vertical collimators feature two jaws: a primary and a secondary jaw. Their MARS geometries were constructed using ROOT TGeo features (https://root.cern.ch). The jaws and surrounding structures were designed to allow parameterized movement, enabling the same model to be used for various operational and accidental conditions.

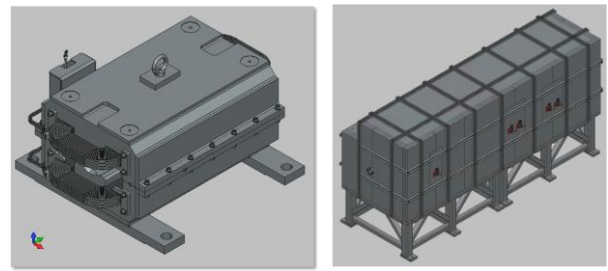

**Fig. 2.** 3D CAD views of the Fast Kicker Dipole (left) and the horizontal PIP-II collimator (right).

Figure 3 shows the operation of the primary jaw. For both the horizontal and vertical collimators, the primary jaw can shift ±25.4 mm from its nominal centered position. A "hard stop" mechanism prevents contact with the vacuum chamber.

---

* Corresponding author: amakovec@fnal.gov

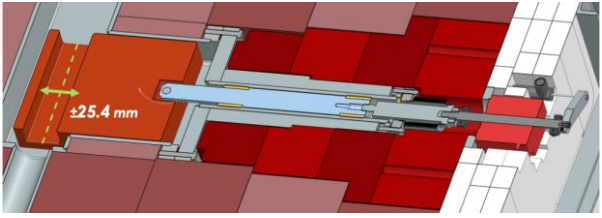

**Fig. 3.** Primary jaw (copper) of the horizontal PIP-II collimator.

Figure 4 presents cross-sectional views of the primary jaw as it moves within the MARS geometry. Notably, the lengths of connected elements adjust accordingly, thanks to the parameterized geometry description.

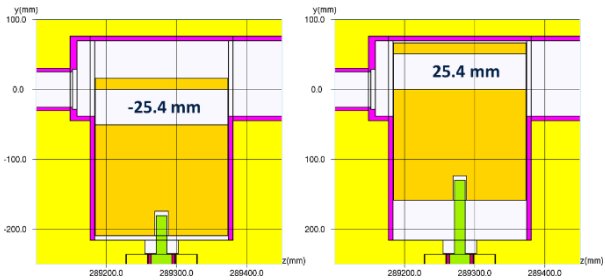

**Fig. 4.** Cross-sectional views showing the motion range of the primary jaw within the MARS geometry.

The secondary jaw operates slightly differently. As shown in Fig. 5, each side of the secondary jaw moves independently, allowing inward movement of up to 25.4 mm from its nominal position, but not outward.

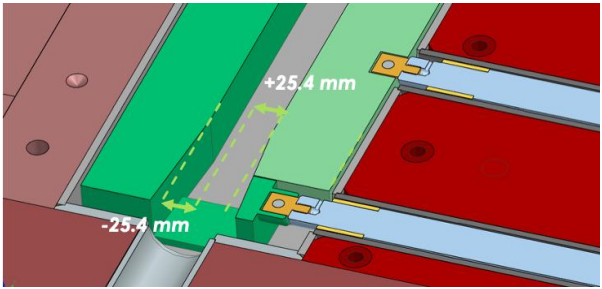

**Fig. 5.** Secondary jaw (stainless steel, SS-304) of the horizontal PIP-II collimator.

Figure 6 illustrates this movement in 2D cross-sectional views, showing the secondary jaw in both fully open and fully closed positions.

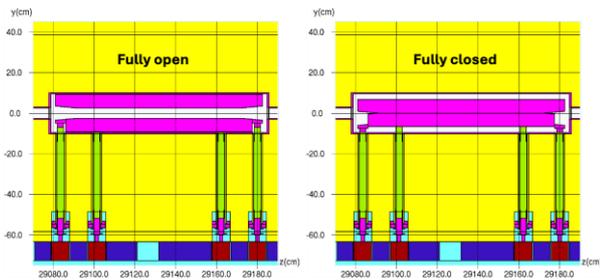

**Fig. 6.** Vertical collimator: secondary jaw shown in fully open (left) and fully closed (right) positions.

The two magnets were implemented using a similar approach. All components (collimators and magnets) were defined in the MARS geometry via ROOT TGeo, enabling repeated placement along the beamline without duplicating geometry definitions.

## 3 Application of Importance-Based Splitting Technique

Figure 7 shows the dose rate map from a MARS simulation for an accident scenario in which the full LINAC beam is lost at the first dipole in the LINAC tunnel for a duration of 3 seconds. The resulting prompt dose rate is scored using a Cartesian grid of $100 \times 1 \times 100$ bins, covering the area shown.

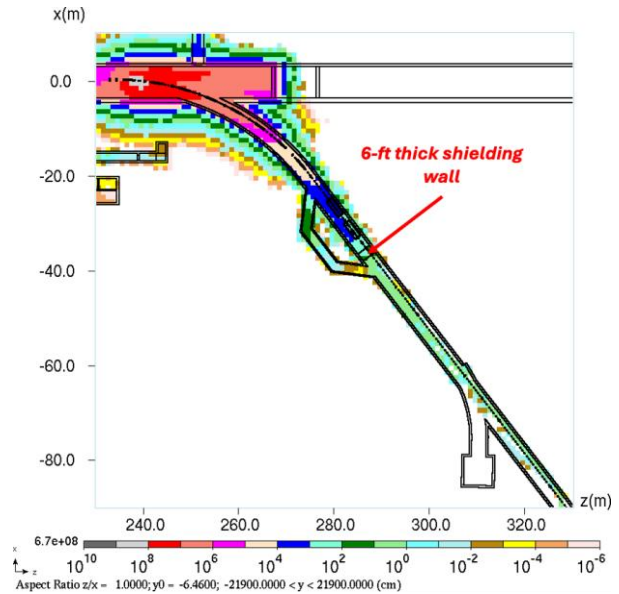

**Fig. 7.** Dose rate map from MARS simulations without branching for the PIP-II LINAC shielding assessment.

No branching technique was applied in this simulation. Despite the long computational time and nearly $10^8$ particle histories, statistical accuracy in the BTL beamline – particularly downstream of the 6-foot shielding wall – is limited, as highlighted in the plot. Beyond this wall, some scoring cells contain no data, indicating that much more computation would be needed. However, due to the complexity of the geometry and the distance from the beam loss point, achieving sufficient statistical accuracy without a branching technique required running the simulation on 1,000 CPU cores for a month, totaling approximately 83 CPU years.

This analysis still allowed us to draw some important conclusions. From the 1D projection of the dose rate map along the BTL beamline and the bypass tunnel, we concluded that the shielding wall reduces radiation levels by more than two orders of magnitude. Furthermore, the bypass tunnel is well-designed to further reduce stray radiation beyond the levels achieved by the shielding wall (see Fig. 8).

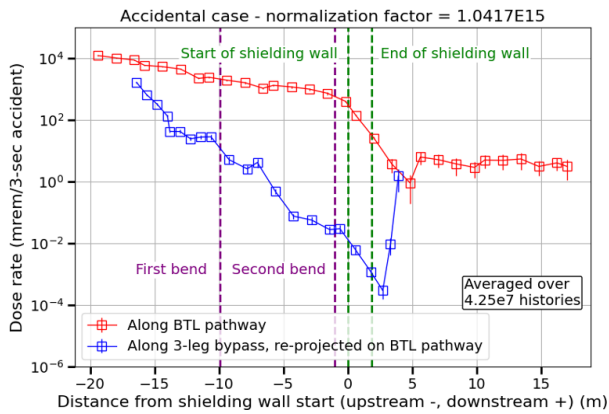

**Fig. 8.** 1D projection of the dose rate map along the BTL beamline and the bypass tunnel.

At the request of the Environment, Safety, and Health (ESH) group, we added several high-resolution detector planes across the tunnel, consisting of cubic foot-sized air cells acting as detectors. These planes allow us to capture radiation field details on a spatial scale comparable to real-world dose monitors. However, implementing them greatly increases computational demands. We estimated that achieving acceptable statistical accuracy under this setup would require over two years of runtime on 1,000 CPU cores.

To reduce runtime while preserving statistical accuracy, we implemented an importance-based particle splitting technique (also known as branching technique) in the MARS code. This method combines particle splitting and the Russian Roulette algorithm, guided by predefined importance values assigned to regions of interest.

Figures 9 and 10 compare dose rate maps generated without and with the branching technique, respectively. Although the comparison is not exact – since Figure 9 shows a 1 GeV accident scenario and Figure 10 an 800 MeV case with lower intensity – meaningful conclusions can still be drawn. The importance-based branching technique improved simulation efficiency by more than a factor of 100.

Moreover, the simulation with branching populated all scoring cells with meaningful data in less than 2 CPU years, while the simulation without branching produced only sparse results after 83 CPU years (see Fig. 7).

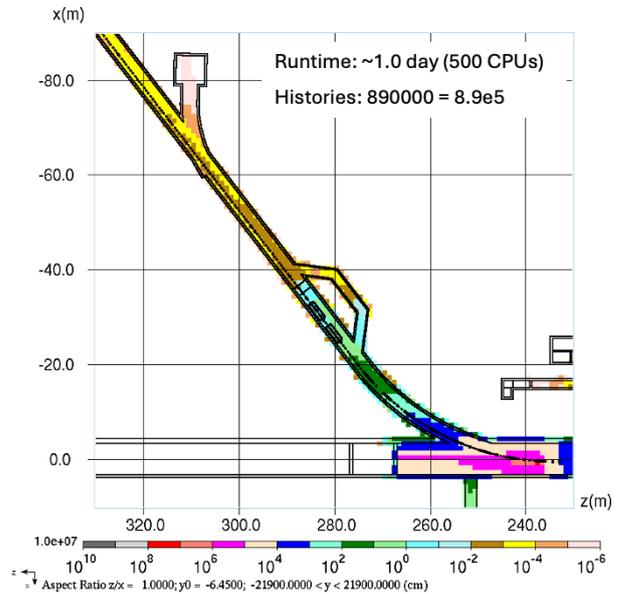

**Fig. 10.** Dose rate map with importance-based branching applied.

## 4 Auxiliary Graphical Tools for Data Processing and Visualization

To support the development of a more user-friendly Graphical User Interface (GUI) for the MARS code, several auxiliary tools were created to improve workflow efficiency and streamline post-processing tasks.

The new GUI is developed in Python, using the PySide6 and Matplotlib libraries, among others. Its main window is designed to allow users to easily switch between different functional areas, such as the Source, Geometry, and Estimator workbenches.

As part of this development, several specialized applications were built. One of these is the MTUPLE Grid Visualizer, a tool for visualizing detector plane results from MTUPLE simulation output files. It offers a user-friendly interface where users can define and manage .ini configuration files for various detector grid settings, select input files, choose output quantities for visualization, and save/load plotting parameters.

The application requires input such as axis limits, the number of bins, and region numbers for grid boundaries (see Fig. 11). Based on these, it automatically constructs the full set of region numbers and loads the corresponding simulation data. This data is displayed on the right side of the window (not shown in Fig. 11). Users can also apply custom normalization factors and choose between continuous or discrete color scales for flexible result visualization.

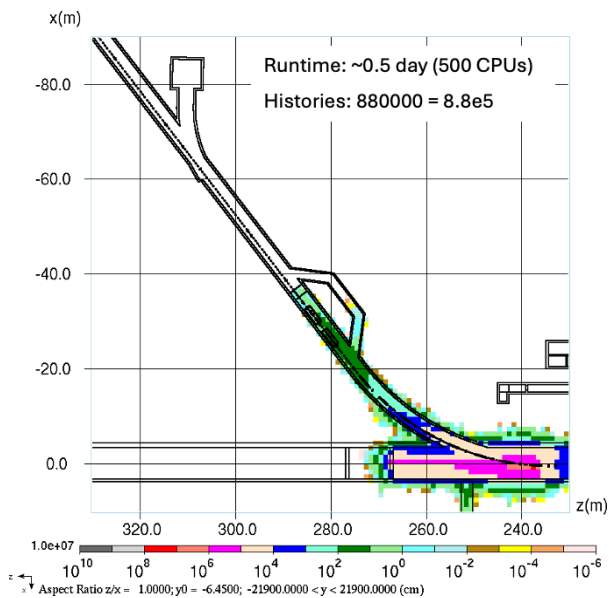

**Fig. 9.** Dose rate map without importance-based branching.

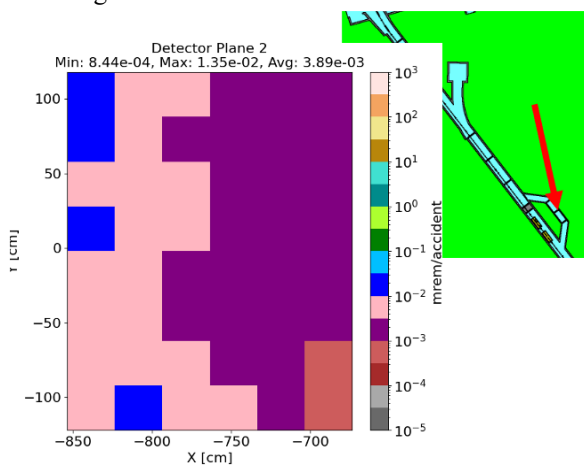

**Fig. 11.** Parameter input panel of the MTUPLE Grid Visualizer. Users can define grid settings, region boundaries, axis parameters, and file selection for data loading and visualization.

Figure 12 shows the application in action, displaying the dose rate map for the middle of the second leg of the bypass tunnel – an output used in the PIP-II LINAC shielding assessment.

**Fi g. 12.** Dose rate map of the middle section of the bypass tunnel, visualized using the MTUPLE Grid Visualizer as part of the PIP-II LINAC shielding assessment.

## 5 Conclusion

The PIP-II Linac Shielding Assessment project at Fermilab necessitated rigorous radiation shielding optimization. Through our comprehensive update of the geometry model and the incorporation of high-resolution detector planes, we have obtained crucial radiation field data essential for this optimization. The development of our importance-based branching code has dramatically reduced simulation runtimes while maintaining statistical integrity, allowing us to efficiently model complex radiation environments. Furthermore, the creation of the MTUPLE Grid Visualizer tool enhances our ability to analyze and interpret detector plane scoring results, making the data more accessible. Overall, our methodologies and tools have significantly improved the efficiency and accuracy of radiation dose assessments for the project.

This work was produced by FermiForward Discovery Group, LLC under Contract No. 89243024CSC000002 with the U.S. Department of Energy, Office of Science, Office of High Energy Physics. Publisher acknowledges the U.S. Government license to provide public access under the DOE Public Access Plan